\newcommand{\simnot}{\mathord{\sim}}
\documentclass[10pt,conference]{IEEEtran}
\IEEEoverridecommandlockouts
\usepackage{cite}
\usepackage{amsmath,amssymb,amsfonts,bbm}

\usepackage{algpseudocode}
\usepackage{graphicx}
\usepackage[caption=false, font=footnotesize]{subfig}
\usepackage{balance}
\usepackage{textcomp}
\usepackage{xcolor}
\usepackage[linesnumbered,ruled,vlined]{algorithm2e}

\SetKwInput{KwInput}{Input}                
\SetKwInput{KwOutput}{Output}              

\def\BibTeX{{\rm B\kern-.05em{\sc i\kern-.025em b}\kern-.08em
    T\kern-.1667em\lower.7ex\hbox{E}\kern-.125emX}}

\begin{document}

\title{Ultra-High Reliability by Predictive Interference Management Using Extreme Value Theory}

%

\author{\IEEEauthorblockN{Fateme Salehi\IEEEauthorrefmark{1}, Aamir Mahmood\IEEEauthorrefmark{1}, Sinem Coleri\IEEEauthorrefmark{2}, Mikael Gidlund\IEEEauthorrefmark{1}}
\IEEEauthorblockA{\IEEEauthorrefmark{1}Department of Computer and Electrical Engineering,  Mid Sweden University, 85170 Sundsvall, Sweden}
\IEEEauthorblockA{\IEEEauthorrefmark{2}Department of Electrical and Electronics Engineering, Koc University, 34450 Istanbul, Turkiye}
Email: \{fateme.salehi, aamir.mahmood, mikael.gidlund\}@miun.se, scoleri@ku.edu.tr
}

\maketitle

\begin{abstract}
Ultra-reliable low-latency communications (URLLC) require innovative approaches to modeling channel and interference dynamics, extending beyond traditional average estimates to encompass entire statistical distributions, including rare and extreme events that challenge achieving ultra-reliability performance regions.
In this paper, we propose a risk-sensitive approach based on extreme value theory (EVT) to predict the signal-to-interference-plus-noise ratio (SINR) for efficient resource allocation in URLLC systems. We employ EVT to estimate the statistics of rare and extreme interference values, and kernel density estimation (KDE) to model the distribution of non-extreme events. Using a mixture model, we develop an interference prediction algorithm based on quantile prediction, introducing a confidence level parameter to balance reliability and resource usage. While accounting for the risk sensitivity of interference estimates, the prediction outcome is then used for appropriate resource allocation of a URLLC transmission under link outage constraints. 
Simulation results demonstrate that the proposed method outperforms the state-of-the-art first-order discrete-time Markov chain (DTMC) approach by reducing outage rates up to 100-fold, achieving target outage probabilities as low as \(10^{-7}\). Simultaneously, it minimizes radio resource usage \(\simnot15 \%\) compared to DTMC, while remaining only \(\simnot20 \%\) above the optimal case with perfect interference knowledge, resulting in significantly higher prediction accuracy. 
Additionally, the method is sample-efficient, able to predict interference effectively with minimal training data.
\end{abstract}


\section{Introduction}
Ultra-reliable low-latency communications (URLLC), a key feature of fifth-generation (5G) and future wireless networks, is designed to provide extremely reliable and low-latency services for mission-critical applications, such as autonomous driving, industrial automation, and remote healthcare. URLLC aims to achieve a latency of less than 1\,ms with a reliability of 99.999 (5-nines). However, advanced industrial applications, such as real-time motion control, require even higher reliability—exceeding 7-nines—and sub-millisecond latency, pushing the limits of 5G’s capabilities \cite{extremeURLLC}. To address these requirements, link adaptation (LA) is essential to dynamically adjust transmission parameters in response to fluctuating channel and interference conditions. 

Predictive resource allocation enhances LA by proactively allocating resources based on the anticipated signal-to-interference-plus-noise ratio (SINR), reducing the need for retransmissions and the risk of outages \cite{QoS_prediction, NB_IoT, KalmanARMA, SaP}. In  \cite{QoS_prediction}, the authors propose a quality of service (QoS) prediction-based radio resource management (RRM) scheme utilizing a predictive proportional fairness resource allocation algorithm to ensure a minimum guaranteed user capacity over the prediction horizon. In \cite{NB_IoT}, a framework for RRM of NB-IoT is proposed that leverages cooperative interference prediction and flexible duplexing techniques. The core idea is to enable cooperation between neighboring base stations (BSs) by exchanging anticipated user scheduling information through backhaul or X-interface before actual scheduling, allowing the interference level during transmission to be predicted more accurately. A recursive predictor that estimates future interference values by a steady-state Kalman filtering is presented in \cite{KalmanARMA}. The predictor’s parameterization is performed offline by translating the autocorrelation of interference into an autoregressive moving average (ARMA) representation. Reference \cite{SaP} proposes a medium access control (MAC) scheme for cognitive radio networks whereby each secondary transmitter (TX) predicts the interference level at the receiver (RX) based on the sensed interference at the TX. The prediction is quantified in terms of spatial interference correlation between the two locations.
However, none of these studies addresses the strict reliability and delay requirements of URLLC systems. 

The existing LA algorithms for URLLC are based on the derivation of the tail statistics of interference distribution from a large number of data samples \cite{KDE,DTMC, 2nd_DTMC}. In \cite{KDE}, a kernel density estimation (KDE) based algorithm is proposed to enhance LA for URLLC. The predicted interference is used to choose a proper modulation and coding scheme to meet a certain one-shot transmission bit error rate (BER) target. In \cite{DTMC}, the interference variation is modeled as a discrete state space discrete-time Markov chain (DTMC). The state transition probability matrix is used to estimate the interference and allocate blocklength accordingly. Reference \cite{2nd_DTMC}, similar to \cite{DTMC}, uses DTMC to model the interference variations. However, it exploits the second-order Markov chain to capture the time correlation among the observed samples, which means that future interference is predicted based on current and past observations. 

However, none of the above methods effectively balance the achievement of the target block error rate (BLER) and outage probability with the constraints of limited data in rapidly changing dynamics.
Achieving ultra-high reliability in URLLC demands a fundamentally different approach from conventional wireless strategies. This shift involves moving from an average-based system design to an extreme value theory (EVT) based statistical framework that specifically models and analyzes tail behaviors \cite{tailURLLC, evt_1, evt_2}. By accurately predicting extreme conditions, resources can be proactively allocated to ensure that the stringent latency and reliability requirements of URLLC are consistently met. Reference \cite{prediction_evt} proposes a sample-efficient framework for predicting rare event statistics by leveraging the spatial dependency between channel measurements from different locations. The framework integrates radio maps with non-parametric models and EVT within a Bayesian framework to estimate channel statistics for rare events. 

\begin{figure}[t]
    \centering
        \includegraphics[width=1\columnwidth,trim={1.5cm 0cm 0cm 0cm},clip]{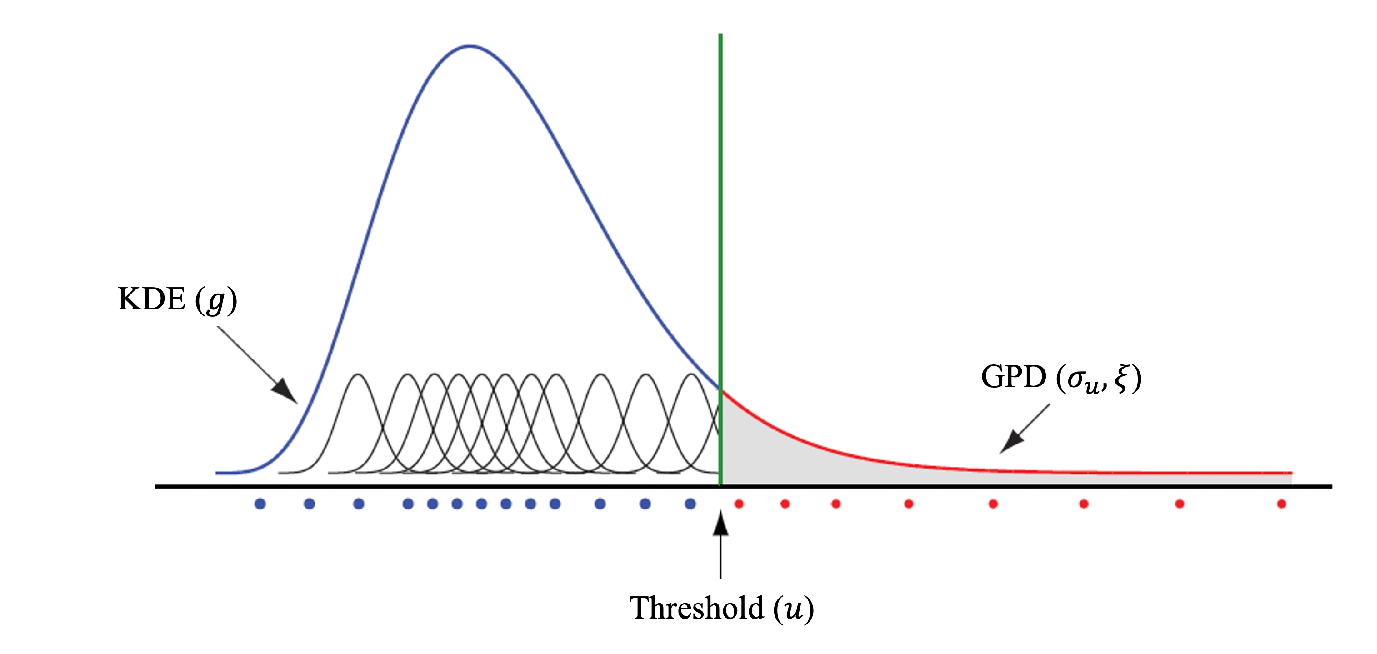}
            \caption{Illustration of the mixture-based interference distribution modeling approach combining kernel density estimation (KDE) and generalized Pareto distribution (GPD). The blue dots and red dots represent data points used for density estimation, respectively in the bulk region and in the tail region.}
            \label{fig.model}
\end{figure}

In this paper, we propose an algorithm based on EVT to predict extreme and rare interference events using limited data samples and incorporate these predictions into resource allocation strategies. The statistics of interference distribution are characterized by a mixture model, where the tail of the interference distribution is estimated using EVT, and the bulk distribution is modeled using the KDE method, as shown in  Fig. \ref{fig.model}. To predict the interference values using the current observation, similarly to \cite{KDE,DTMC,2nd_DTMC}, we apply quantile prediction by considering a confidence level parameter to ensure the reliability requirement.
Eventually, the predicted interference is mitigated through efficient resource allocation. The proposed method is compared with the first-order Markov-based interference prediction in \cite{DTMC} as a baseline scheme.

The rest of the paper is organized as follows. Section \ref{sec.sys} outlines the system model. Section \ref{sec.method} describes the proposed interference prediction method and the corresponding resource allocation strategy for ultra-reliable communication. Section \ref{sec.res} presents the numerical results and compares the performance of the proposed method with baseline schemes. Finally, Section \ref{sec.con} provides the conclusions of the paper.

\section{System Model}\label{sec.sys}
We consider a wireless network consisting of a desired URLLC link and $N$ independent interferers. URLLC transmissions occur over mini-slots of duration $\simnot0.1$\,ms. Therefore, time is discretized into mini-slots denoted by $t$. At a given mini-slot $t$, the $k${th} transmitter is either idle or transmitting with power $p_k(t)$. The probability of a new transmission of each interferer is denoted by $\mu$, which is defined as the activation factor. The message duration $\zeta$ is the same for all transmissions. A single-antenna Rayleigh block-fading channel model is adopted, given that the channel coherence time in a typical wireless environment is much larger than the mini-slot duration. The aggregate interference at the location of the desired user at time $t$ is the sum of the receiving powers from all interfering nodes and is given as
\begin{equation}
\label{eq.sumInterf}
    I(t)=\sum_{k=1}^N p_k(t) \hslash_k^2(t) \delta_k(t)\ , 
\end{equation}
where $\hslash_k(t)$ is the channel coefficient between the $k$th interferer and the desired user, and $\delta_k(t)$ is a Bernoulli random variable capturing the traffic intensity given by
\begin{equation}
\label{eq.bernoul}
    \delta_k(t) = \left\{ \begin{array}{l}
1,~{\textrm{ if node }}k{\textrm{ is transmitting,}}\\
0,~{\textrm{ otherwise.}}
\end{array} \right. 
\end{equation}
which depends on the parameters $\mu$ and $\zeta$. 
Hereinafter, we represent $I(t)$ as $I^t$ for notational simplicity.

According to \eqref{eq.sumInterf}, two factors cause the correlation of interference: time-varying traffic and time-varying channel. The first one is reflected by random access of interferers described earlier by parameters $\mu$ and $\zeta$ and captured in $\delta_k(t)$, and the second one is denoted by $\hslash_k(t)$. 
As the transmission occurs over mini-slots in URLLC, the channel can also be the origin of correlation. In this case, since the channel coherence time covers several transmission intervals, the interference from a specific transmitter is correlated over time.

The desired transmitter sends a short packet of $b$ bits with a target outage $\epsilon$. The transmitter estimates the SINR and then allocates the required resources accordingly. The goal is to estimate the interference and consequently, SINR as accurately as possible to ensure efficient resource allocation. URLLC aims to achieve high reliability and low latency simultaneously, typically by meeting a reliability target under a specific latency constraint. While retransmissions enhance reliability, they also increase latency. In this study, we focus on the lower bound of achieved reliability, where guaranteeing successful transmission within the stringent constraints is critical. To accurately assess the worst-case performance, we assume that only one-shot transmission is allowed, as this eliminates the possibility of retransmissions, which would introduce delays and additional overhead.

\section{The Proposed Interference Prediction and Resource Allocation Method}\label{sec.method}

\subsection{Interference Prediction}
In this paper, we adopt a mixture model-based approach to predict the interference, where the entire distribution function is approximated as a combination of \textit{bulk distribution} for non-extreme events and \textit{tail distribution} for extreme events based on a considered threshold level. Extreme and non-extreme events are often physically caused by different underlying processes, which implies there is little information in the bulk distribution for describing the tail behavior. Available mixture models have been broadly classified by the type of bulk distribution models into parametric (fully/semi-parametric) and non-parametric models \cite{mixture}. To characterize the interference distribution, we propose a non-parametric model based on a kernel density estimator for the bulk distribution, combined with a generalized Pareto distribution (GPD) for the tail model as shown in Fig.\ref{fig.model}.

The interference is predicted based on the previous observation, capturing the time correlation among samples, similarly to \cite{DTMC, KDE}. The underlying assumption is that the signal-to-noise ratio (SNR) of the desired transmission and the interference power in the previous transmissions are perfectly estimated by the BS. In a practical system, these two quantities are estimated by the user based on pilot signals received from the BS and are then fed back to the BS via the channel quality indicator (CQI) \cite{KDE}. In this respect, note that CQI feedback mechanisms of current cellular systems already provide precise knowledge of the SINR in previous transmissions.

As the first step, the initially observed interference space $\mathcal{I}$, measured over consecutive time instants, is divided into the state space $\mathcal{L}\overset{\Delta}{=}\{\mathcal{S}_1,\dots,\mathcal{S}_L\}$ with $L$ unequal levels, where interference values in the range $[\mathbb{I}_{l-1},\mathbb{I}_{l})$ are assigned to state $\mathcal{S}_l$. The boundaries could be chosen such that each state $\mathcal{S}_l$ corresponds to a specific quantile range of the observed interference. We utilize quantile-based clustering for the sake of risk management for URLLC. It ensures that extreme interference values, which are critical for URLLC, are captured in specific states with sufficient resolution.
Next, we group the interference space $\mathcal{I}$ into $L$ disjoint clusters called conditional interference space $\mathcal{I}_l$, such that $\cup_{l=1}^L \mathcal{I}_l = \mathcal{I}$ and
\begin{equation}
    \mathcal{I}_l=\{I^{t+1}|\mathbbm{1}_{\mathcal{S}_l}(I^{t})\} \ ,
\end{equation}
where $\mathbbm{1}_\mathcal{A}(x)$ is the indicator function, which equals 1 if $x\in\mathcal{A}$ and 0 otherwise.

We aim to predict the interference and consequently the SINR for LA, and thus reducing the probability of transmission failures. We apply the quantile prediction by considering a certain percentile of the conditional interference distribution. An outage will occur if the predicted interference is lower than the actual value.
Hence, the interference power is estimated so that the predicted interference, $\hat{I}^{t+1}$, is greater than the actual interference, $I^{t+1}$, with a certain probability called the confidence level parameter $\eta$, i.e., 
\begin{equation}
\label{eq.MQ}
    P \left(I^{t+1} \leq \hat{I}^{t+1}|\mathbbm{1}_{\mathcal{S}_l}(I^{t})\right) \geq \eta \ . 
\end{equation}

In order to compute $\hat{I}^{t+1}$, we need to estimate the conditional probability density function (PDF) of $I^{t+1}$ given previous observation $I^t \in \mathcal{S}_l$, i.e., $f(I^{t+1}|\mathbbm{1}_{\mathcal{S}_l}(I^t))$. From the conditional interference space $\mathcal{I}_l$, we can approximate it with $f_{\mathcal{I}_l}(I^{t+1})$. Then the left side of \eqref{eq.MQ} becomes
\begin{equation}
\label{eq.int}
\begin{split}
     P \left(I^{t+1} \leq \hat{I}^{t+1}|\mathbbm{1}_{\mathcal{S}_l}(I^{t})\right) & = \int_0^{\hat{I}^{t+1}} f(\phi|\mathbbm{1}_{\mathcal{S}_l}(I^t))\, d\phi \\ & = \int_0^{\hat{I}^{t+1}} f_{\mathcal{I}_l}(\phi)\, d\phi \ .  
\end{split}
\end{equation}

\subsubsection{GPD for the Tail Distribution}
EVT provides a robust framework for analyzing the statistics of extreme events that rarely happen by modeling the probabilistic distribution of the values exceeding a given threshold using the GPD \cite{EVT_book}. To determine the statistics of the interference tail distribution of each cluster $\mathcal{I}_l$, we first consider a rare event threshold based on a fixed percentile. 
To simplify notation, we consider a generic random variable $X=[x_1,x_2,\dots,x_N]$ with i.i.d. samples and PDF of $h(x)$ for extreme events, i.e. the events exceeding the threshold $u$, and PDF of $g(x)$ for non-extreme events. The mixture model is given by
\begin{equation}
    f(x) = \left\{ \begin{array}{l}
g(x),~{x < u,}\\
h(x),~{x \geq u.}
\end{array} \right.
\end{equation}

The probabilistic distribution of the values exceeding the threshold, i.e., $P\left(X - u \leq y | X > u \right)$, is approximated by the GPD with cumulative distribution function (CDF) given by
\begin{equation}
    H_u\left(y\right)=1-\left(1+\frac{\xi y}{\sigma_u}\right)^{-1 / \xi} ,
\end{equation}
where $y$ represents a non-negative value denoting the exceedance above threshold $u$, $\sigma_u > 0$ is the scale parameter and $\xi \in \mathbb{R}$ is the shape parameter \cite{EVT_book}.

Denote by $H(x)$ the tail distribution of $X$, i.e., $H(x)=P(X \leq x)$ for $x > u$. Given the GPD parameters for the threshold exceedance, it follows that
\begin{equation}
\label{F_tail}
    H\left(x\right)=\left(1+\frac{\xi(x-u)}{\sigma_u}\right)^{-1 / \xi} \cdot \rho_u \ ,
\end{equation}
where $\rho_u=P(X \geq u)$. The extreme quantile $X_{\varepsilon} > u$, that is expected to be exceeded with probability $\varepsilon$ on average follows directly from $H_X^{-1}(\varepsilon)$ (with $H_X$ the CDF of extreme values of $X$ and $\varepsilon$ close to zero $\varepsilon = 1 - \eta$) and \eqref{F_tail} as
\begin{equation}
\label{eq.extreme}
    X_\varepsilon=u+\frac{\sigma_u}{\xi}\left(\left(\frac{\rho_u}{\varepsilon}\right)^{\xi}-1\right) .
\end{equation}
Given threshold $u$, the GPD parameters are estimated using the maximum likelihood estimator (MLE).

\subsubsection{KDE for the Bulk Distribution}
We adopt the KDE approach to compute the PDF of the conditional interference space $\mathcal{I}_l$ below the exceedance threshold. The samples are filtered with the kernel function $K$, from which we obtain the PDF. Using KDE, we approximate the distribution $g(x)$ for $x < u$ as
\begin{equation}
    g(x)=\frac{1}{N_{\text{b}}} \sum_{n=1}^ {N_{\text{b}}} K\left(\frac{x_n-x}{T}\right),
\end{equation}
where $K$ is the kernel function and \(N_{\text{b}}\) is the number of samples in the bulk region. The kernel bandwidth, $T$, is a parameter of the kernel function which must be suitably selected.
We consider the Gaussian kernel function, i.e.,
\begin{equation}
    K\left(\frac{x_n-x}{T}\right)=\frac{1}{\sqrt{2 \pi T}} e^ {-(x_n-x)^2/(2 T)}.
\end{equation}
Bandwidth $T$ can be obtained by minimizing the asymptotic mean integrated squared error \cite{KDE_diffusion, KDE}.
 
 \subsection{Resource Allocation}
In this section, we use the interference prediction for LA to meet the target outage of URLLC users. Let $\hat{\gamma}=\sigma/(1+\hat{I})$ be the predicted SINR at the user location, where $\sigma$ is the SNR of the desired transmission and $\hat{I}$ is the predicted interference-to-noise ratio (INR) as described earlier. In the finite blocklength regime, the number of $b$ information bits can be transmitted in a blocklength of $M$ channel uses with a decoding error probability of $\epsilon$ in an additive white Gaussian noise (AWGN) channel with SINR $\gamma$, which is given as \cite{polyanskiy}
\begin{equation}
\label{eq.FBL}
    b \approx MC(\gamma)-Q^{-1}(\epsilon)\sqrt{MV(\gamma)} \ , 
\end{equation}
where $C(\gamma)=\log_2(1+\gamma)$ is the Shannon capacity of AWGN channel and $V(\gamma)=(\log_2(e))^2(1-(1+\gamma)^{-2})$ is the channel dispersion, and $Q^{-1}$ is the inverse of the Gaussian Q-function. Using \eqref{eq.FBL}, the blocklength $M$ can be approximated as \cite{chUsage}
 \begin{equation}
\label{eq.chUse}
    M \approx \frac{b}{C(\gamma)}+\frac{Q^{-1}(\epsilon)^2V(\gamma)}{2C(\gamma)^2}\!\left[1\!+\!\sqrt{1+\frac{4bC(\gamma)}{Q^{-1}(\epsilon)^2V(\gamma)}}\right]\!. 
\end{equation} 
The characteristics of our channel model remain the same in each realization and can be assumed as an AWGN channel.

\subsection{Implementation Perspective}
The proposed interference prediction algorithm is low in complexity and requires minimal additional signaling overhead. Constructing the CDF \( f(x) \), where GPD parameters and KDE are fitted, is handled offline during the training phase. At runtime, the only additional step is estimating interference based on the current interference level using either GPD or KDE, depending on the confidence level and exceedance threshold. Interference estimation using EVT is performed via \eqref{eq.extreme}, utilizing GPD parameters \((\sigma_u, \xi)\) and threshold \( u \), while KDE-based estimation uses the CDF according to \eqref{eq.MQ}. 

The algorithm requires the receiver to provide feedback on the experienced interference level alongside the CQI feedback and requires the transmitter to update GPD or KDE parameters after receiving this information. Notably, frequent parameter updates after each observation are unnecessary; instead, periodic or batch updates can be employed to maintain computational efficiency without compromising interference prediction accuracy.

\section{Numerical Results}\label{sec.res}

The performance of the proposed interference prediction method is numerically evaluated against the two baseline schemes: the first-order DTMC estimation proposed in \cite{DTMC} and the genie-aided estimation. The genie-aided interference estimate, in which the transmitter has prior knowledge of the exact interference conditions, is an idealized or impractical benchmark that we consider as an indicator of the optimum performance bound. The main simulation parameters are adopted as in Table~I unless stated otherwise. The simulation results are obtained by averaging over 100 random INRs, each with a training phase of \(10^3\) samples, followed by a runtime phase of \(10^5\) channel realizations.

\begin{table}[!t]
\centering
\caption{Simulation Parameters}
\label{tab.setup}
\begin{center}
\begin{tabular}{ |l|c|l| } 
 \hline
 \textbf{Parameter} & \textbf{Symbol} & \textbf{Value}\\
 \hline
 Mean SNR & $\sigma$ & $20$ dB\\
 \hline
 Mean INR range & & $\mathcal{U}[-10,5]$ dB \\
 \hline
 Packet size & $b$ & $50$ bits \\ 
 \hline
 Number of interferes & $N$ & $5$ \\ 
 \hline
Number of quantile levels & $L$ & $15$ \\ 
 \hline
Exceedance threshold & $u$ & $97$th percentile \\ 
 \hline
 Activation factor & $\mu$ & $[0.4,1]$ \\ 
 \hline
 Message duration & $\zeta$ & $10$ mini-slots \\ 
 \hline
 Filter length for time correlation & & 100 \\
 \hline
 Number of training samples & & $10^3$ \\
 \hline
 Confidence level & $\eta$ & $\{0.95,0.99\}$ \\ 
 \hline
 \end{tabular}
\end{center}
\end{table}

\subsection{Performance Under Different Confidence Levels}
\begin{figure}[t]
    \centering
        \includegraphics[width=1\columnwidth]{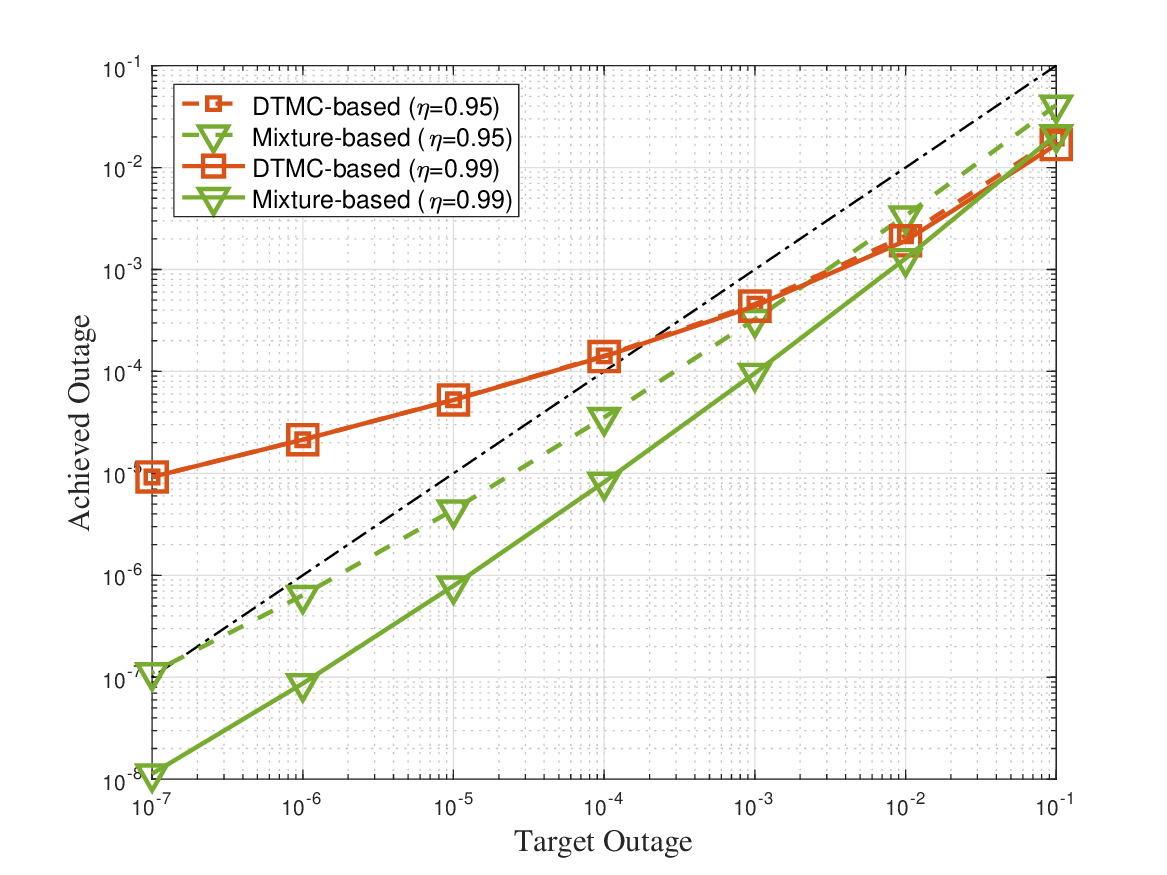}
            \caption{Comparison of achieved outage vs. target outage probability for the proposed mixture-based method against DTMC-based method for different confidence levels $\eta = \{0.95,0.99\}$.}
            \label{fig.unLimit_out}
\end{figure}

\begin{figure}[t]
    \centering
        \includegraphics[width=1\columnwidth]{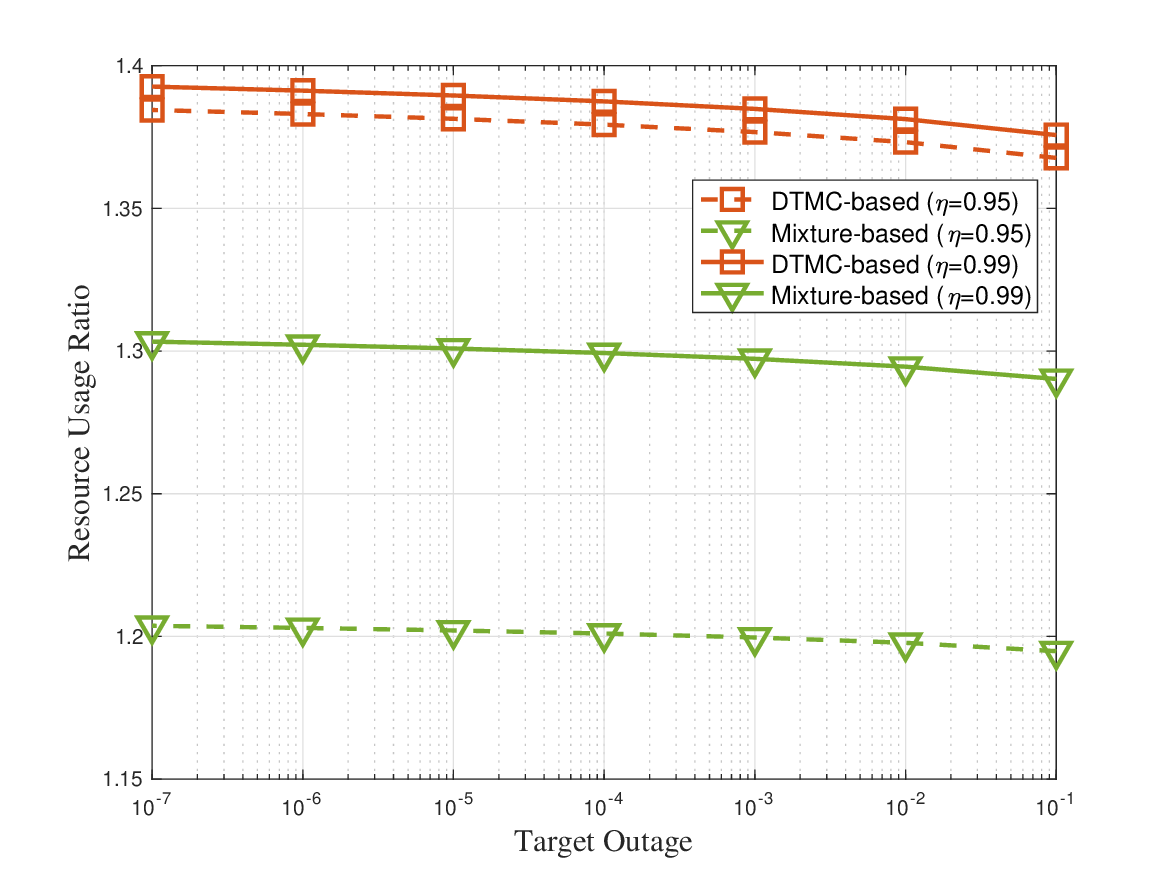}
            \caption{Comparison of resource usage ratio vs. target outage probability for the proposed mixture-based method against DTMC-based method for different confidence levels $\eta = \{0.95,0.99\}$.}
            \label{fig.unLimit_res}
\end{figure}

Fig. \ref{fig.unLimit_out} and Fig. \ref{fig.unLimit_res} compare the performance of the proposed mixture-based interference prediction method against the DTMC-based method for various target outage probabilities from $10^{-1}$ to $10^{-7}$. In Fig. \ref{fig.unLimit_out}, the achieved outage is demonstrated for two different confidence levels ($\eta = 0.95$ and $\eta=0.99$). It is observed that the proposed method achieves a lower outage compared to the DTMC-based method at the same confidence level, implying better accuracy. In Fig. \ref{fig.unLimit_res}, the resource usage ratio of different schemes to the genie-aided estimation is demonstrated for the same confidence levels. The mixture-based method consistently outperforms the DTMC-based method in both reducing achieved outage and minimizing resource usage, regardless of the confidence level. The performance of the mixture method is closer to the genie-aided estimation at $\eta = 0.95$ as it requires blocklength about only $1.2$ times the genie-aided estimation. 
Increasing the confidence level from $\eta = 0.95$ to $\eta = 0.99$ leads to improved performance in achieved outage, which results in higher resource usage. Hence, a more careful balancing is required for the best trade-off between outage performance and resource efficiency for both methods. However, the DTMC-based method exhibits lower sensitivity to the confidence level than the mixture-based method, largely due to its reliance on a discretized interference space and estimation approach.
Note that the mixture-based method at $\eta=0.95$ provides an estimation based on KDE since the confidence level is lower than the exceedance percentile, i.e. $0.97$, but for $\eta=0.99$ the estimation is based on EVT.

\subsection{Performance Under the Resource Limit}
\begin{figure}[t]
    \centering
        \includegraphics[width=1\columnwidth]{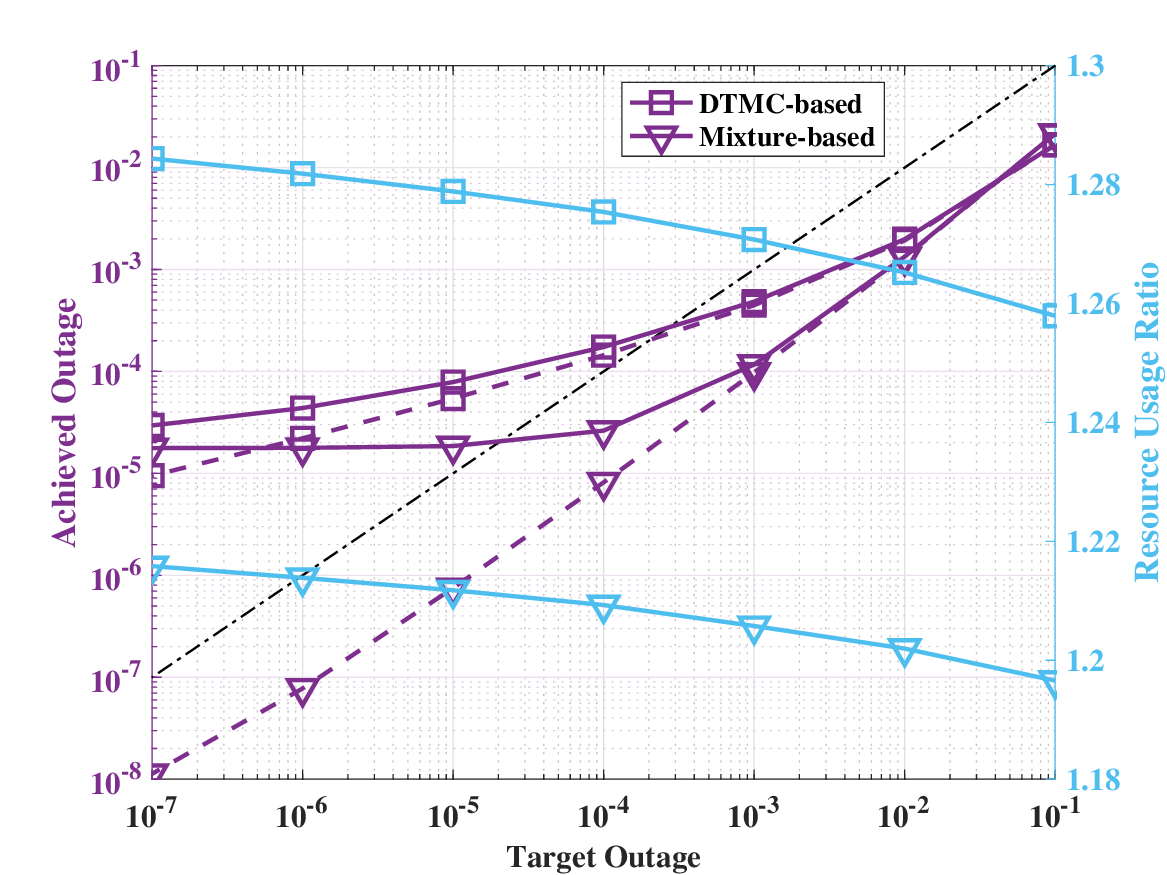}
            \caption{Achieved outage and resource usage ratio vs. target outage probability under blocklength limit $M=10^5$ channel uses and confidence level $\eta=0.99$. The dashed lines indicate the $99.99$th percentile of the achieved outage.}
            \label{fig.limit}
\end{figure}

Owing to random variations in the channel and interference, the received SINR of the desired user may occasionally drop. When this occurs, the required blocklength to meet the target outage increases significantly, potentially making it impractical.
In Fig. \ref{fig.limit}, we compare the performance of the proposed method under the blocklength limit of $M=10^5$ and the confidence level of $\eta=0.99$. The mixture-based method, similar to the DTMC-based method, exhibits a tendency to overshoot the target outage. However, unlike the DTMC-based approach, the $99.99$th percentile line in the mixture-based method lies significantly lower than its average value. This demonstrates that the achieved outage meets the outage probability across all target levels, indicating better performance and more accurate interference control. 
Additionally, Fig. \ref{fig.limit} shows the resource usage ratio of different methods. The mixture-based method requires $\simnot5 \%$ fewer resources than the DTMC-based approach, indicating better efficiency. Thus, the mixture-based method achieves precise interference control with efficient resource usage compared to the DTMC-based approach.

\subsection{Performance Under Different Number of Training Samples}
\begin{figure}[t]
    \centering
        \includegraphics[width=1\columnwidth]{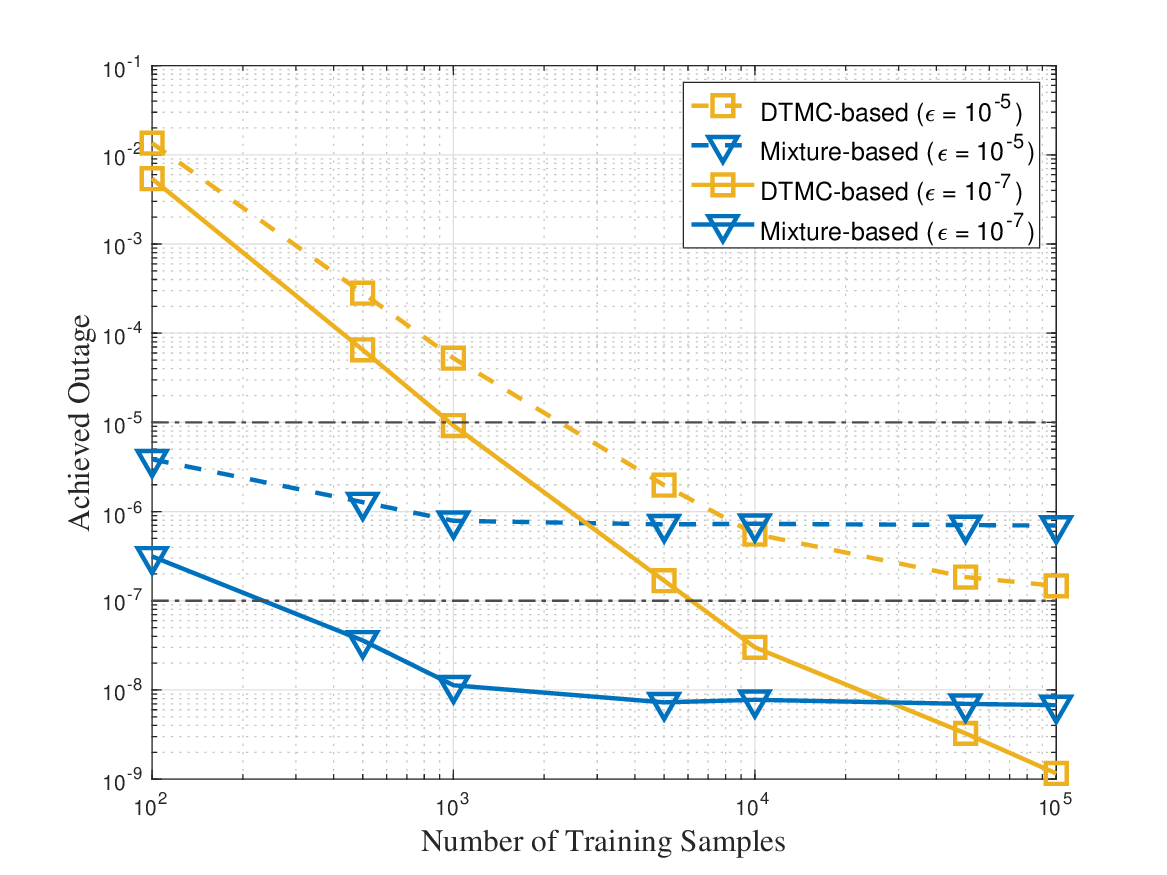}
            \caption{Comparison of the achieved outage vs. the number of training samples for the proposed mixture-based method against DTMC-based method for $\eta=0.99$ and different target outages $\epsilon = \{10^{-5},10^{-7}\}$.}
            \label{fig.samples}
\end{figure}

In Fig. \ref{fig.samples}, we compare the performance of the proposed scheme and the baseline in terms of the number of training samples. The DTMC-based method requires a large number of training samples, for example, about $3000$ samples to meet the desired target outage of $10^{-5}$, whereas the mixture-based method achieves this with only about $100$ samples.
The achieved outage decreases as the number of training samples increases. However, it is important to note that training the DTMC with $10^5$ samples and achieving a lower outage than the mixture-based method does not necessarily indicate better prediction accuracy, as this reduction is attributed to overestimation and increased resource usage. 
The mixture-based method provides a better trade-off between the number of training samples and the achieved outage for both target levels ($\epsilon=10^{-5}$ and $\epsilon=10^{-7}$). The DTMC-based method struggles to achieve similar performance, particularly at the more stringent target ($\epsilon=10^{-7}$), requiring significantly more samples.
The proposed method reaches a stable level at the outage with about $1000$ training samples, indicating higher learning efficiency even with fewer samples compared to the DTMC-based method. Overall, the mixture-based method is more sample-efficient, meaning it can achieve low outage values with fewer training samples, making it a more reliable and efficient choice for interference prediction, especially when limited training data is available.

\section{Conclusion}\label{sec.con}
In this paper, we addressed the challenge of ensuring reliability in ultra-reliable communication through interference prediction. Unlike conventional wireless scenarios, retransmissions are impractical here due to the associated time overhead. Instead, intelligent and risk-sensitive strategies are required to meet the stringent demands for reliability and latency while maintaining high resource efficiency. The primary motivation of this research is to achieve high reliability while minimizing resource wastage during link adaptation (LA) with the constraints of limited data in rapidly changing dynamics. To this end, accurately predicting the signal-to-interference-plus-noise ratio (SINR) is crucial for efficient resource allocation.
Leveraging extreme value theory (EVT), we estimated the statistics of the interference tail and approximated the bulk distribution using the kernel density estimation (KDE) method. Based on the mixture model, we proposed an algorithm to predict interference in the next time slot using quantile prediction and the confidence level parameter to balance reliability and resource usage. We then used the predicted SINR for blocklength allocation.
We compared the proposed interference prediction method with the state-of-the-art first-order discrete-time Markov chain (DTMC) and demonstrated that the proposed mixture-based approach is a promising solution for ultra-high reliability communication with stringent target outage probabilities as low as \(10^{-7}\). Simulation results confirmed that the mixture-based method consistently outperforms the DTMC-based method by both reducing achieved outage and minimizing resource usage, indicating higher prediction accuracy.
Furthermore, we showed that our proposed method effectively predicts interference with a very low number of training samples, ensuring the required reliability. In other words, the proposed mixture-based approach is sample-efficient, making it a more reliable choice for interference prediction, particularly in scenarios with limited training data or rapidly changing system dynamics. Future work could focus on refining the threshold selection strategy for EVT-based solutions and identifying the optimal confidence level parameter for rate allocation.


\bibliographystyle{IEEEtran}
\bibliography{reference.bib}

\end{document}